\documentclass[nofootinbib,preprint,showpacs,preprintnumbers,amsmath,amssymb,natbib]{revtex4}
\usepackage{graphicx}
\usepackage{epsfig}		
\usepackage{dcolumn}
\usepackage{bm}
\usepackage{multirow}
\usepackage{longtable}
\usepackage{array}
\usepackage{booktabs}
\newcommand{\head}[1]{\textnormal{\textbf{#1}}}
\newcommand{\normal}[1]{\multicolumn{1}{l}{#1}}

\def\0{\mbox{\tiny $0$}}
\def\1{\mbox{\tiny $1$}}
\def\2{\mbox{\tiny $2$}}
\def\3{\mbox{\tiny $3$}}
\def\4{\mbox{\tiny $4$}}
\def\5{\mbox{\tiny $5$}}
\def\6{\mbox{\tiny $6$}}
\def\7{\mbox{\tiny $7$}}
\def\8{\mbox{\tiny $8$}}
\def\9{\mbox{\tiny $9$}}

\def\f14{\mbox{\tiny $\frac{1}{4}$}}

\DeclareMathOperator{\sech}{sech}

\begin{document}

\title{Braneworld scenarios from deformed defect chains}

\author{M. Chinaglia}
\email{chinaglia.mariana@gmail.com}
\author{A. E. Bernardini}
\email{alexeb@ufscar.br}
\affiliation{Departamento de F\'{\i}sica, Universidade Federal de S\~ao Carlos, PO Box 676, 13565-905, S\~ao Carlos, SP, Brasil}
\author{Rold\~ao da Rocha}
\email{roldao.rocha@ufabc.edu.br}
\affiliation{Centro de Matem\'atica, Computa\c c\~ao e Cogni\c c\~ao,
Universidade Federal do ABC 09210-580, Santo Andr\'e, SP, Brazil}
\date{\today}

\begin{abstract}

Novel braneworld scenarios supported by warp factors driven by a single extra dimension are obtained from deformed one-dimensional lump-like solutions known {\em a priori}.
Through a novel {\em ansatz}, the internal energy structure, the braneworld warp factor, and the quantum mechanical analogue problem, as well as the associated zero mode solutions, are straightforwardly derived by means of an analytical procedure.
The results allow one to identify thick brane solutions that support internal structures and that can hold the ($3+1$)-dimensional gravity. 
\end{abstract}

\date{\today}
\maketitle

\section{Introduction}

Braneworld scenarios driven by extra dimensions \cite{modelosmundobrana1,modelosmundobrana3,modelosmundobrana4}
have been identified through several strategies involving analytical procedures.
In these scenarios, the observable universe is supported by a hypersurface with $3+1$ dimensions, the brane, inserted in a bulk with $(3+s)+1$ dimensions, where $s$ is the number of extra dimensions.
Paradigmatically, the extra dimension can be very large or even infinitely extended, and thereat particles and fields are supposed to be constrained by the brane properties, whereas  gravity can evade as to permeate the entire bulk. Matter localization is possible due to a warp factor parameter driving the metric that folds the brane through an approach that can explain, for instance, the hierarchy properties \cite{hierarquia1,RandallSundrum} beyond the Standard Model.

Despite its challenging proposal, the search for beyond the Standard Model solutions of descriptions of the Universe has become an arduous task.
The brane scenarios considered here, and their matter localization profiles exhibited in some cases, are related to generic solutions of an effective action driven by a real scalar field, $\zeta$, coupled to $4+1$-dimensional gravity, given by
\begin{equation}
S = \int dx^5\, \sqrt{\det{g_{AB}}}\left(-\frac{1}{4}R + \frac{1}{2}g_{AB}\partial^A \zeta\partial^B \zeta - \mathcal{V}(\zeta)\right),
\label{00000}
\end{equation}
where $R$ is the scalar curvature, $g_{AB}$ denotes the metric tensor, with $A$ and $B$ running from $0$ to $4$, and
which gives rise to several strategies for investigating analytical braneworld scenarios.

Real scalar field theories coupled to gravity produce thick domain walls often related to integrable models through which analytical solutions of gravitating defect structures admit an internal structure.
In this case, potentials driven by a single real scalar field support BPS-like solutions \cite{BookB} of first-order differential equations that allow for generating braneworld scenarios through the insertion of deformed topological defects \cite{Bas01,Bas05B,topologicalorigin}.
In one-dimensional scalar field theories resumed by a scalar field $\chi(x) \neq \zeta$ (from Eq.~(\ref{00000})), deformed $\lambda \chi^4$ and sine-Gordon models \cite{lumpalex, meulump}, with field potentials respectively given by $V(\chi)=(1/2)(1-\chi^2)^2$ and $V(\chi)=1-\cos(\chi)$, can thus trigger off such novel braneworld scenarios driven by $\zeta$.
In order to guarantee the existence of defect solutions, one imposes that the potential $V(\chi)$ generates a set of critical points, $\verb"{" \bar{\chi}_1,\ldots,\bar{\chi}_n\verb"}"$, such that $(dV/d\chi\vert_{\chi=\bar{\chi}_i}=0)$ as well as $V(\bar{\chi}_i)=0$, with $i=1,2,\ldots,n$.

Deformed potentials exhibiting either a single or multiple minima engender two types of defects with different topologies. When it has at least two degenerate minima, it generates two new vacua corresponding to the classical field configurations, $\chi =\pm\bar{\chi}$. In one spatial dimension the space boundaries consist in two critical points of the potential, i. e. those located at $x = -\infty$ and $x = +\infty$. This spatial boundary topology is mirrored by the vacua configuration space topology of the fields, which also consists in two points, $\chi =-\bar{\chi}$ and $\chi =+\bar{\chi}$. On the other hand, when the potential has one single minimum, $\chi=\bar{\chi}$, it leads to $\lim\chi_{x\rightarrow\pm \infty}=\bar{\chi}$. Consequently, a trivial map carries both boundary spatial points in the same field value ($-\bar{\chi}$ or $+\bar{\chi}$), which defines lump-like structures. Likewise, for degenerate minima, the identity map carries $x = -\infty$ into $\chi =-\bar{\chi}$ and $x = +\infty$ into $\chi =+\bar{\chi}$, which defines kink-like structures.

In order to guarantee the model integrability as well as matter localization, several lump-like defects can be obtained from deformation chains (as those from Refs.~ \cite{lumpalex} and \cite{meulump}).
Departing from a primitive defect, $\chi$, which triggers of a deformation chain -  which might be identified by $\lambda \chi^{4}$ or eventually a sine-Gordon kink solutions - one is able to engender $N\hspace{-.1 cm}+\hspace{-.1 cm}2$ - cyclic deformation chains through the use of a recurrence chain rule constrained by hyperbolic and trigonometric elementary relations (see the Appendix) \cite{lumpalex,meulump}.

In particular, the deformation chain results can be used to build up novel warp factor structures of braneworlds.
To pursue such a proposal, this paper is organized as follows.
In Section II, one reports about the braneworld equations of motion in $4 + 1$ dimensions, as well as to the general procedure for computing metric perturbations.
In Section III, one introduces the {\em ansatz} for obtaining warp factors from one-dimensional lump-like solutions that support a set of novel braneworld scenarios also described in this section.
Finally, our conclusions are shown in Section IV.

\section{Braneworld first order formalism and metric perturbations}

Let one considers the action from Eq.~(\ref{00000}) for a $4+1$ dimensional model driven by the RS2 metric \cite{RandallSundrum},
\begin{equation}
ds^2=e^{2A}\eta_{\mu\nu}dx^{\mu}dx^{\nu}-dy^2,
\label{intervaloinvariante}
\end{equation}
where $\eta_{\mu\nu} = diag\{+1,\,-1,\,-1,\,-1\}$, with $\mu,\nu=0,\, 1,\, 2,\, 3$, and where the extra dimension appears in the warp factor function, $e^{2A(y)}$, that localizes matter around $y=0$.
Assuming a dependency of $\zeta$ exclusively on $y$, the equation of motion for the scalar field  arising from the action (\ref{00000}) is given by
\begin{equation}
\frac{d^2\zeta}{dy^2} + 4 \frac{d A}{dy} \frac{d \zeta}{dy} - \frac{d}{d\zeta}\mathcal{V}(\zeta) = 0.
\label{004}
\end{equation}
By varying the action, $S$, with respect to the metric $g_{\mu\nu}$, one obtains
\begin{equation}
\frac{3}{2}\frac{d^2 A}{dy^2} = -  \left(\frac{d\zeta}{dy}\right)^2,
\label{00005}
\end{equation}
and after some mathematical manipulations, one has
\begin{equation}
3 \left(\frac{d A}{dy}\right)^2 = \frac{1}{2} \left(\frac{d\zeta}{dy}\right)^2 - \mathcal{V}(\zeta)
\label{0006}
\end{equation}
which are the Einstein equations rewritten in a convenient form.

The scalar field potential is used to be written in terms of a {\em superpotential}, $w$, in a particular form given by
\begin{equation}
\mathcal{V}(\zeta) = \frac{1}{8}\left(\frac{dw}{d\zeta}\right)^2 - \frac{1}{3} w^2,
\label{potencialcomW}
\end{equation}
which has been largely considered in the literature \cite{DeWolfe,Gremm,Erlich,Afonso,Sasakura}.

In order to demote from second to first order field equations, a superpotential $w(\zeta)$ is introduced as to set \cite{Gremm,Erlich,Brane},
\begin{equation}
\mathcal{V}(\zeta)=\frac{1}{8} \bigg( \frac{dw(\zeta)}{d\zeta} \bigg)^2-\frac{1}{3} w(\zeta)^2.
\label{potencialcomW}
\end{equation}
such that the above introduced field equations can be rewritten as first order equations,
\begin{equation}
\label{philinha}
\frac{d \zeta}{dy}=\frac{1}{2}\frac{dw(\zeta)}{d\zeta},
\end{equation}
\begin{equation}
\frac{d A}{dy}=-\frac{1}{3}w(\zeta).
\end{equation}

In order to examine the distribution of matter inside the brane, one recovers the energy density derived from (\ref{00000}) \cite{Arias,Barbosa,Borninfield},
\begin{equation}
T^{\zeta}_{00} = T_0^{\zeta\,0}g_{00} = \bigg[ \bigg( \frac{d\zeta}{dy} \bigg)^2-3 \bigg( \frac{dA}{dy} \bigg)^2 \bigg]e^{2A(y)},
\label{em}
\end{equation}
and, in the same scope, a quantum mechanical analogue to examine the stability of metric fluctuations is motivated by the interest of finding a resonance spectrum like those exhibited by {\em volcano} potentials.

Gravitational wave quantum analogue equations can be straightforwardly derived through tensor metric perturbations in  Eq.~(\ref{intervaloinvariante}), as to have
\begin{equation}
ds^2 = e^{2 A (y)} \,\left( \eta_{\mu\nu} + h_{\mu\nu}\right)\,dx^{\mu}\,dx^{\nu} - d y^2,
\label{001F}
\end{equation}
where a factor $e^{2 A (y)}$ is extracted from the fluctuation term as to simplify subsequent equations, with $h_{\mu\nu} \equiv  h_{\mu\nu} (x,y)$ in the form of transverse and traceless tensor perturbations, for which one obtains
\begin{equation}
\left(\frac{d^2~ }{dy^2} + 4\frac{d A}{dy}\frac{d~ }{dy} - e^{-2A(y)}\partial_\mu\partial^{\mu}\right) h_{\mu\nu}(x,y) = 0,
\label{001G}
\end{equation}
for the linearized gravity decoupled from the scalar field.
If the zero mode solution of the resonance spectrum is normalizable - which can be translated into some kind of localization - it is possible to localize the four-dimensional gravity inside the brane.

Considering the convenient change of variable,
\begin{equation}
dz=e^{-A(y)}dy,
\end{equation}
and denoting $h_{\mu\nu}\sim e^{i k^{\nu} x_{\nu}} e^{-3A/2}H_{\mu\nu}(z)$, with $k$ arbitrary, one arrives at the one-dimensional Schr\"odinger equation,
\begin{equation}
\label{sch}
-\frac{d^2H_{\mu\nu}}{dz^2}+U(z)H_{\mu\nu}=k^2\,H_{\mu\nu},
\end{equation}
where $U(z)$ is the quantum mechanical analogue potential given by
\begin{equation}
U(z)=\frac{3}{2}A^{\prime\prime}(z)+\frac{9}{4}A^{\prime 2}(z).
\label{QMpotential}
\end{equation}
Through Eq.~(\ref{sch}), one recovers the Hamiltonian form given by
\begin{equation}
H= \bigg( -\frac{d}{dz}-\frac{3}{2}A^{\prime} \bigg) \bigg( \frac{d}{dz} -\frac{3}{2}A^{\prime} \bigg),
\label{hamiltoniana}
\end{equation}
with $k$ real and $k^2\geq0$, as to not exhibit unstable modes.
By means of Eq.~(\ref{hamiltoniana}), one finds the zero mode solution ($k=0$),
\begin{equation}
H_{\mu\nu}(z)=N_{\mu\nu}e^{3A(z)/2},
\label{modozero}
\end{equation}
where $N_{\mu\nu}$ is the normalization constant.

Finally, a short comment related to reducing second order nonlinear equations to a system of firs order nonlinear equations is pertinent, in the sense discussed in the end of Section V of the Ref.~\cite{DeWolfe}, where it has been stablished that such order reduction must be followed by the correct absorption of the related integration constants, implicitly considered along our calculations.

The above equations summarize the simplified mathematical description of the brane scenarios with one extra dimension, such that the warp factor calculation procedure can be setup in the next section.

\section{Analytical {\em ansatz} for the warp factors}\label{metodo}

Assuming that localized energy structures on the brane demand for warp factors delineated by a lump-like behavior, which guarantees the model integrability and matter localization, one identification of a localized behavior described by an arbitrary {\em bell-shaped} function, $\alpha(y)$, can be performed through
\begin{equation}
e^{2A(y)}=\alpha(y),
\end{equation}
with $\alpha(y)$ known \emph{a priori}. Thereat, $A(y)$ is given straightforwardly by
\begin{equation}
A(y)=\frac{1}{2}\ln[{\alpha}(y)],
\label{A}
\end{equation}
as to have
\begin{equation}
w(y)=-3\frac{dA}{dy} = -\frac{3}{2}\frac{1}{{\alpha}} \frac{d{\alpha}}{dy}.
\end{equation}
A plethora of kink an lump-like analytical defects were released in the Ref. \cite{lumpalex} generated from cyclic deformation chains (see the Appendix) triggered by the solutions derived from a $\lambda \chi^4$ model (besides other three deformed models).
Several of those lump-like solutions can be inserted into Eq.~(\ref{A}) as to generate braneworld scenarios \cite{Bas01,Afonso,Bas05B,Bas05,Bas04,Bas02}.

\subsection{Obtaining deformed defects}

Departing from some previously known analytical profile for a defect, $\chi$, a deformation procedure \cite{Bas01} sets that the driving potential can be rewritten as
\begin{equation}
V(\chi)= \tilde{V}[\tilde{\chi}(\chi)]\left(\frac{d\tilde{\chi}}{d\chi}\right)^{-2},
\end{equation}
where the deformation function,  $\tilde{\chi}_{\chi} \equiv d\tilde{\chi}/d\chi$, relates minimal points from the primitive defect, $\chi$, to minimal points from deformed solutions, $\tilde{\chi}$, in such a way that,
\begin{equation}
\tilde{\chi}(y)\equiv\tilde{\chi}[\chi(y)],
\label{def}
\end{equation}
where $V$ and $\tilde{V}$ are respectively primitive and deformed potentials.
As for triggering a cyclic deformation chain, it has been shown that the deformation procedure can exhibit a topological mass constraint  \cite{lumpalex} given by
\begin{equation}
\sum_n m_{\tilde{\chi}_n}=m_{\chi}.
\end{equation}
Thereat the summation of the $n$ deformed defect topological masses is equal to the primitive defect mass, in a kind of cyclic chain of one-dimensional defect deformation.
Cyclic chains of kink and lump-like deformed defects have been built through the above procedure by means of deformation functions, $\tilde{\chi}_{\chi}$, exhibiting either hyperbolic or trigonometric behaviors (see the Appendix) \cite{lumpalex,meulump}.

Since deforming chains with three or four connected defects ($3$-cyclic, for $N+2 = 3$, and $4$-cyclic, for $N+2 = 4$, deformations) have been considered, the above indexed label $\tilde{\chi}$ is given in certain generic sense: the primitive kink solution is always labeled by $\chi$, while different Greek letters ($\alpha=\psi,\phi$ and $\varphi$) are used to index the deformed defects.

The primitive kink solutions that shall have triggered off the deformation defect generation depicted from $N+2$ - cyclic chains (a $3$-cyclic one, with $N=1$, and a $4$-cyclic one with $N=2$) from Ref. \cite{lumpalex} have been given by
\begin{eqnarray}
\chi_1(y)&=& \tanh(y) ~ \qquad\mbox{($\lambda\chi^4$ model)}, \\
\chi_2(y)&=& \sech(y), \\
\chi_3(y)&=& \ln[1+\sech(y)^2], \\
\chi_4(y)&=& 2(2+y^2)^{-1},
\end{eqnarray}
from which, deformed defects identified by $\tilde{\chi}$ into Eq.~(\ref{def}) could be obtained and identified with $\alpha$ from Eq.~(\ref{A}), $\alpha\sim\tilde{\chi}$ as to support warp factors that generate consistent braneworld scenarios driven by a scalar field $\zeta$ as set by the action from (\ref{00000}).
That is, the deformed defects depicted from $3$ and $4$-cyclic deformation chains from Refs. \cite{lumpalex} that shall be used to construct novel braneworld scenarios parameterized by $\alpha(y)\sim\tilde{\chi}(y)$ are summarized in Table \ref{tabela}, which describes a plenty of deformed lump-like solutions (and the respective correspondence with Refs. \cite{lumpalex,meulump}). They shall be analytically described in the sections where they are required.
\begin{table}[h!]
\centering
\begin{tabular}{@{}*3l@{}}
\toprule[1.5pt]
& \normal{\head{3-cyclic}} & \normal{\head{4-cyclic}}
 \\
  \cmidrule(lr){2-3}
  \hline
  \multirow{3}{3cm}{Hyperbolic} &  $\psi(\chi_1(y))$ & $\phi(\chi_1(y))$  \\
&  &  $\varphi(\chi_2(y))$  \\
&  &  $\varphi(\chi_3(y))$  \\
&  &  $\varphi(\chi_4(y))$    \\
  \cmidrule(lr){2-3}  \hline
  {Trigonometric} & $\psi(\chi_1(y))$ & $\varphi(\chi_2(y))$  \\
\cmidrule(lr){1-1}
\bottomrule[1.5pt]
\end{tabular}
\caption{Greek letters for the lump-like deformed defects derived with hyperbolic and trigonometric functions in \cite{lumpalex} ($2^{nd}$ and $3^{rd}$ columns). The primitive defects listed here trigger off the deforming chain that are described in the following sequence from paragraphs $1$ to $5$.}
\label{tabela}
\end{table}

According to Eq.~(\ref{A}), the localization profiles described by such deformed solutions, $\alpha(y)\sim\tilde{\chi}(y)$, apart from a normalization constant, shall reproduce the localization profiles of the warp factors, $e^{2A(y)}$, for each one of those novel braneworld scenarios.

In the following, one shall identify the results obtained for all the scenarios summarized by Table \ref{tabela}, from which the warp factor is identified by Eq.~(\ref{A}), and used as an input function for computing the energy density, $T_{00}(y)$ (c. f. Eq.~(\ref{em})), and the corresponding QM potential, $U(z(y))$ (c. f. Eq.~(\ref{QMpotential})).

\subsubsection{Braneworld for a lump derived from the $3-$cyclic hyperbolic chain in a $\lambda \chi^4$ model}

\begin{figure}[h!]
\centering
\includegraphics[scale=0.37]{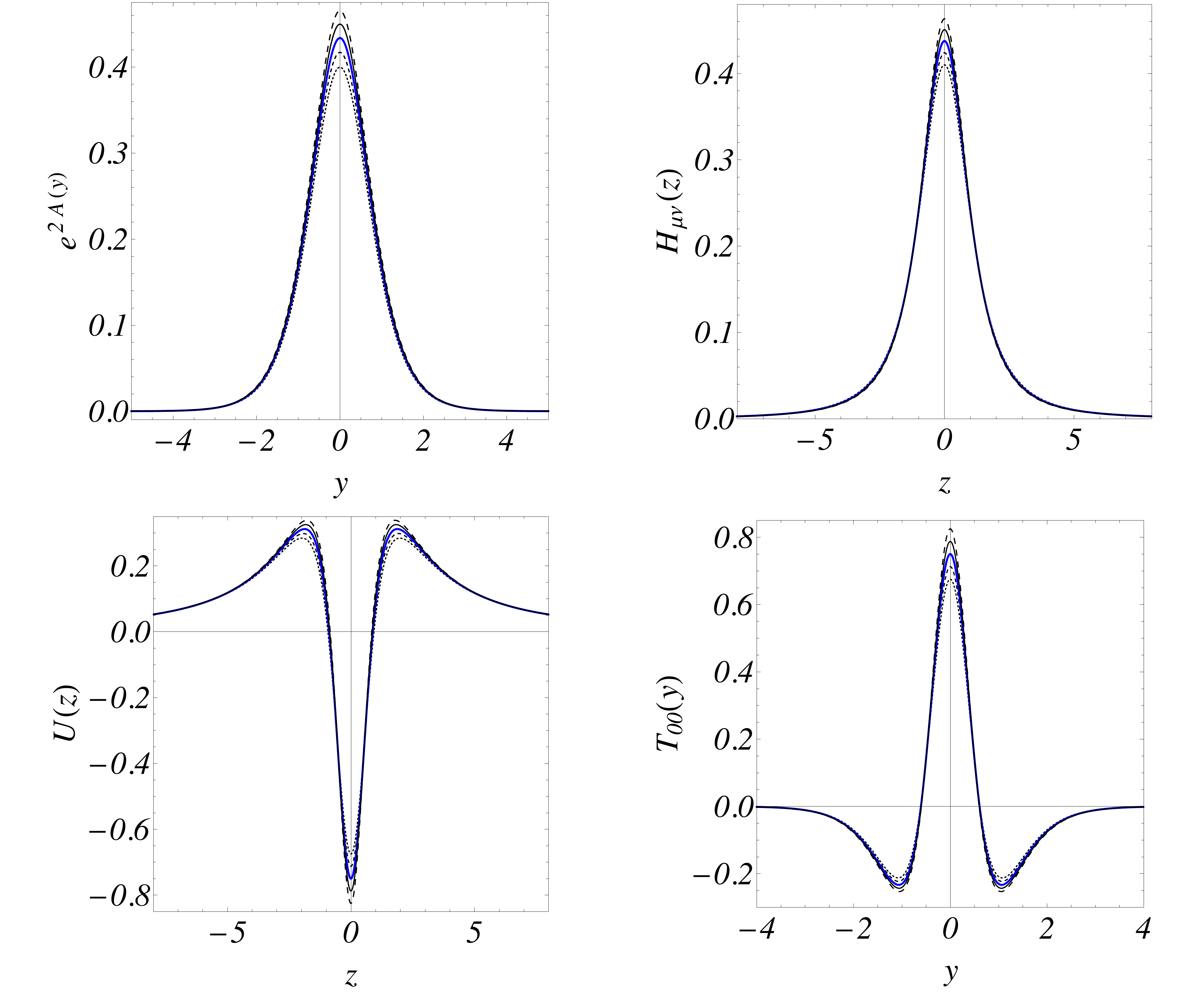}
\caption{Model derived from $\psi(\chi_1)$ respective to the $3-$cyclic hyperbolic chain. Warp factor ($1^{st}$ column) and zero mode solution ($2^{nd}$ column) are presented in the first line ; QM potential as a function of $z$ ($1^{st}$ column) and the energy density ($2^{nd}$ column) are presented in the second line. Blue line represents the solution for $n = 1$, noting that $n=1-2\epsilon;1-\epsilon;1;1+\epsilon;1+2\epsilon$, with $\epsilon=0.05$. Black lines range from dotted to dashed as $n$ increases.}
\label{lambda1}
\end{figure}

The first scenario that hereon considered is driven by the \emph{ansatz}:
\begin{equation}
\psi(\chi_1(y))=\frac{-\ln[\cosh[n \tanh(y)] \sech(n)]}{n},
\end{equation}
is thus summarized by

\begin{equation}
A(y) = \frac{1}{2} \ln \bigg[\frac{-\log[\cosh[n \tanh(y)] \sech(n)]}{n} \bigg],
\end{equation}

\begin{equation}
T_{00}(y) = \frac{3}{4} \bigg[ n \sech(y)^4 \sech[n\tanh(y)]^2 -2 \sech(y)^2\tanh(y)\tanh[n \tanh(y)] \bigg], \\
\end{equation}

\begin{eqnarray}
U(y) &=&\frac{3}{16} \sech(y)^2 \bigg[8 \tanh(y) \tanh[n \tanh(y)] \nonumber\\ && \quad\quad + n \sech(y)^2 \bigg(-4 \sech[n \tanh(y)]^2-\frac{\tanh[n \tanh(y)]^2}{\log[\cosh[n \tanh(y)] \sech(n)]}\bigg) \bigg],
\end{eqnarray}
from which the zero mode is straightforwardly calculated through Eq.~(\ref{modozero}).
The results are depicted in Fig.~\ref{lambda1}, where lump solutions depend on a free parameter $n$, for which one notices that $n=1-2\epsilon;1-\epsilon;1;1+\epsilon;1+2\epsilon$, with $\epsilon=0.05$.

\subsubsection{Braneworld for a lump derived from the $3-$cyclic trigonometric chain in a $\lambda \chi^4$ model}

\begin{figure}[h!]
\centering
\includegraphics[scale=0.4]{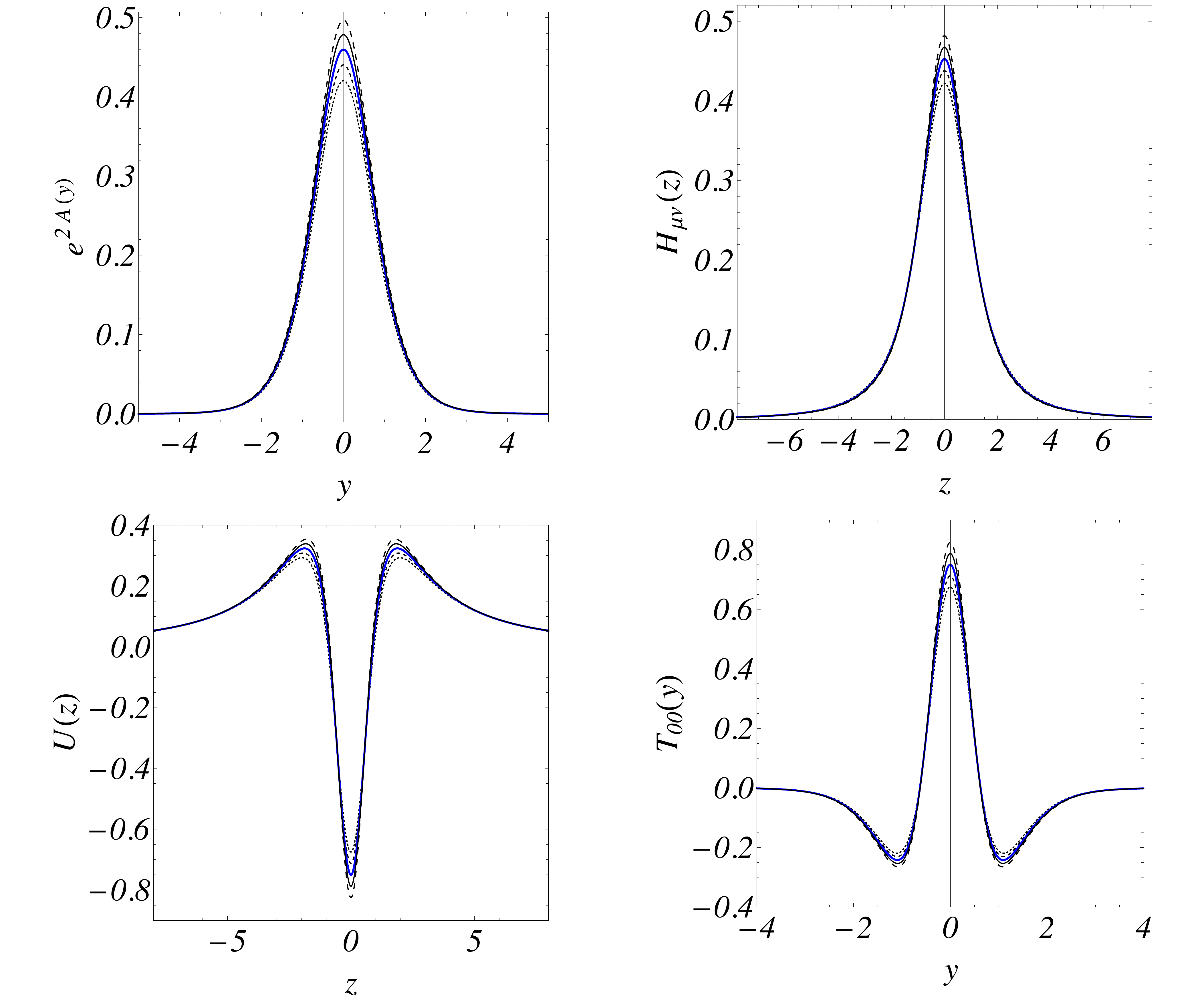}
\caption{Model derived from $\psi(\chi_1)$ respective to the $3-$cyclic trigonometric chain. Warp factor ($1^{st}$ column) and zero mode solution ($2^{nd}$ column) are presented in the first line; QM potential as a function of $z$ ($1^{st}$ column) and the energy density ($2^{nd}$ column) are presented in the second line. Blue line represents the solution for $n = 1$ while black lines range from dotted to dashed as $n$ increases (as for the same parameters of Fig.~\ref{lambda1}).}
\label{lambda2}
\end{figure}

A similar braneworld scenario obtained now for a deformed solution from a trigonometric chain with the \emph{ansatz}:
\begin{equation}
\psi(\chi_1(y))=\frac{\cos[n \tanh(y)] - \cos(n)}{n},
\end{equation}
is thus summarized by
\begin{eqnarray}
A(y) &=& \frac{1}{2} \ln[\frac{-\cos(n) + \cos[n \tanh(y)]}{n}], \\
T_{00}(y) &=& \frac{3}{4} \bigg[ n \cos[n \tanh(y)] \sech(y)^4 - 2 \sech(y)^2 \sin[n \tanh(y)] \tanh(y) \bigg],
\end{eqnarray}
\begin{eqnarray}
U(y) &=& \frac{3 n (3 - 8 \cos(n) \cos[n \tanh(y)] + 5 \cos[2 n \tanh(y)]) \sech(y)^4}{
 32 (\cos(n) - \cos[n \tanh(y)])} \nonumber\\
 &&\quad\quad\quad\quad\quad\quad\quad\quad\quad +\frac{3}{2} \sech(y)^2 \sin[n \tanh(y)] \tanh(y).
\end{eqnarray}
for which the results are depicted in Fig.~\ref{lambda2}.

\subsubsection{Braneworld for a lump derived from the $4-$cyclic hyperbolic chain in a $\lambda \chi^4$ model}

\begin{figure}[h!]
\centering
\includegraphics[scale=0.4]{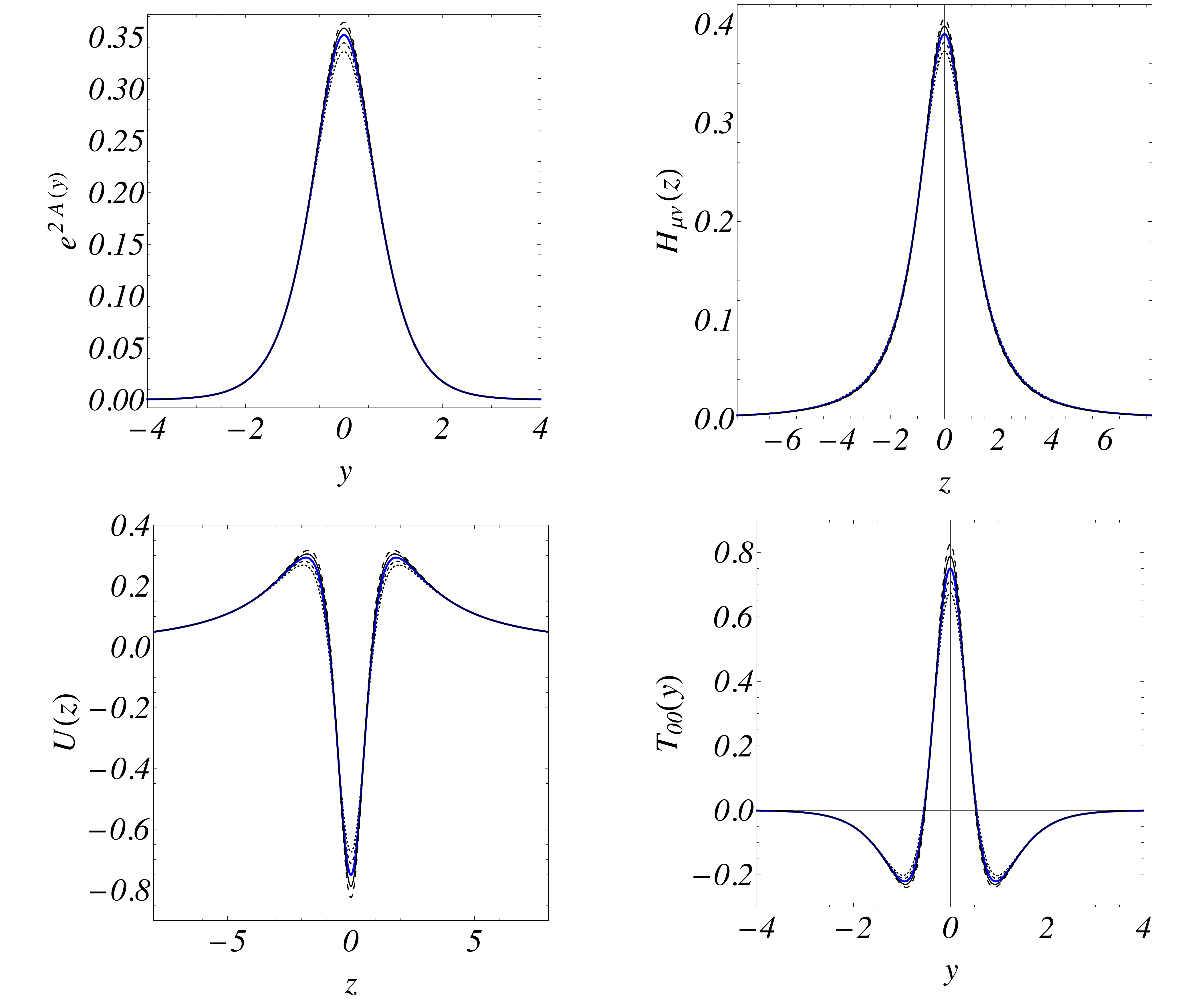}
\caption{Model derived from $\phi(\chi_1)$ respective to the $4-$cyclic hyperbolic chain. Warp factor ($1^{st}$ column) and zero mode solution ($2^{nd}$ column) are presented in the first line; QM potential as a function of $z$ ($1^{st}$ column) and the energy density ($2^{nd}$ column) are presented in the second line. Blue line represents the solution for $n = 1$ while black lines range from dotted to dashed as $n$ increases (as for the same parameters of Fig.~\ref{lambda1}).}
\label{lambda3}
\end{figure}

\begin{figure}[h!]
\centering
\includegraphics[scale=0.4]{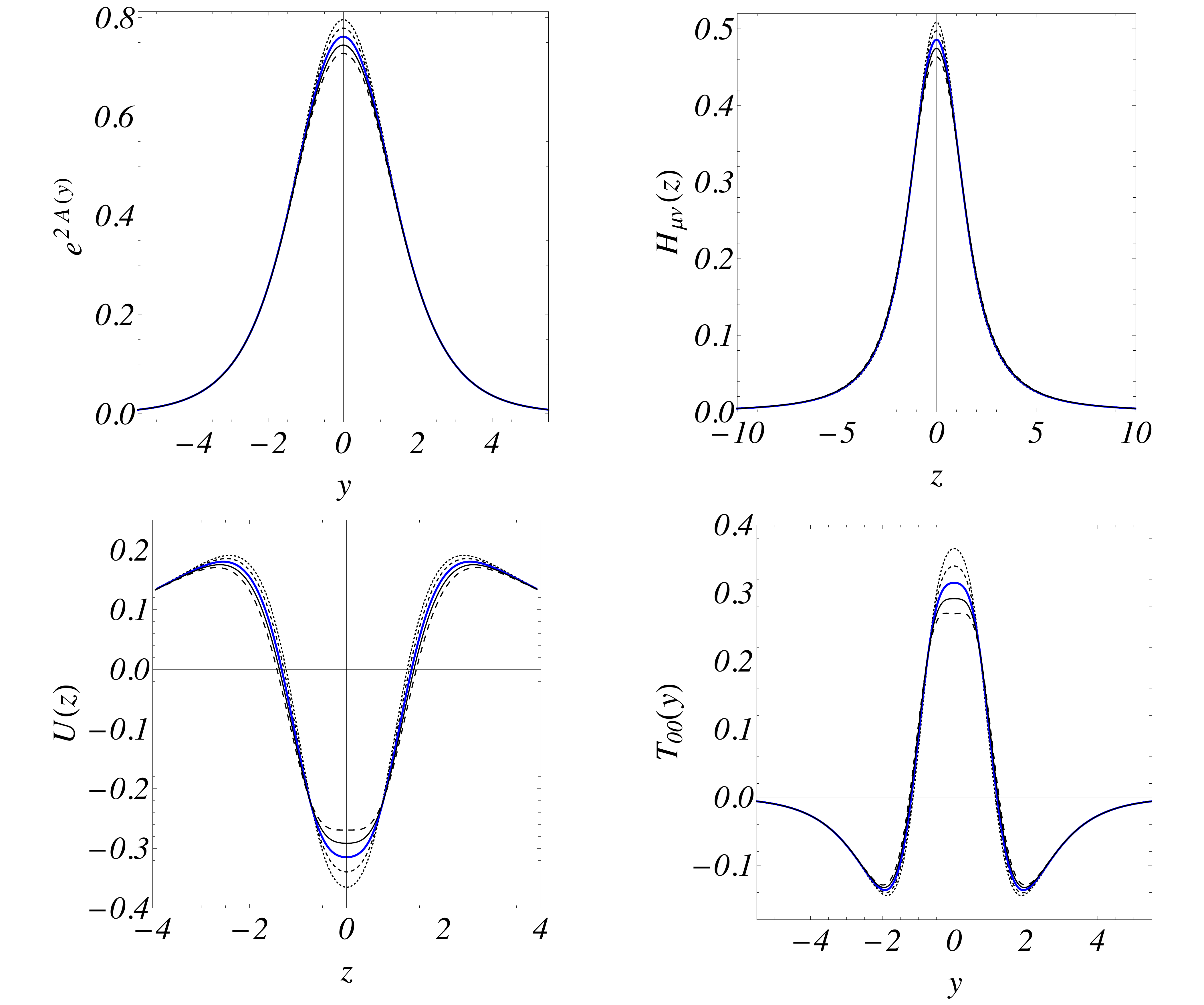}
\caption{Model derived from $\varphi(\chi_2)$ respective to the $4-$cyclic hyperbolic chain. Warp factor ($1^{st}$ column) and zero mode solution ($2^{nd}$ column) are presented in the first line; QM potential as a function of $z$ ($1^{st}$ column) and the energy density ($2^{nd}$ column) are presented in the second line. Blue line represents the solution for $n = 1$ while black lines range from dotted to dashed as $n$ increases (as for the same parameters of Fig.~\ref{lambda1}).}
\label{lambda4.1}
\end{figure}

\begin{figure}[h!]
\centering
\includegraphics[scale=0.4]{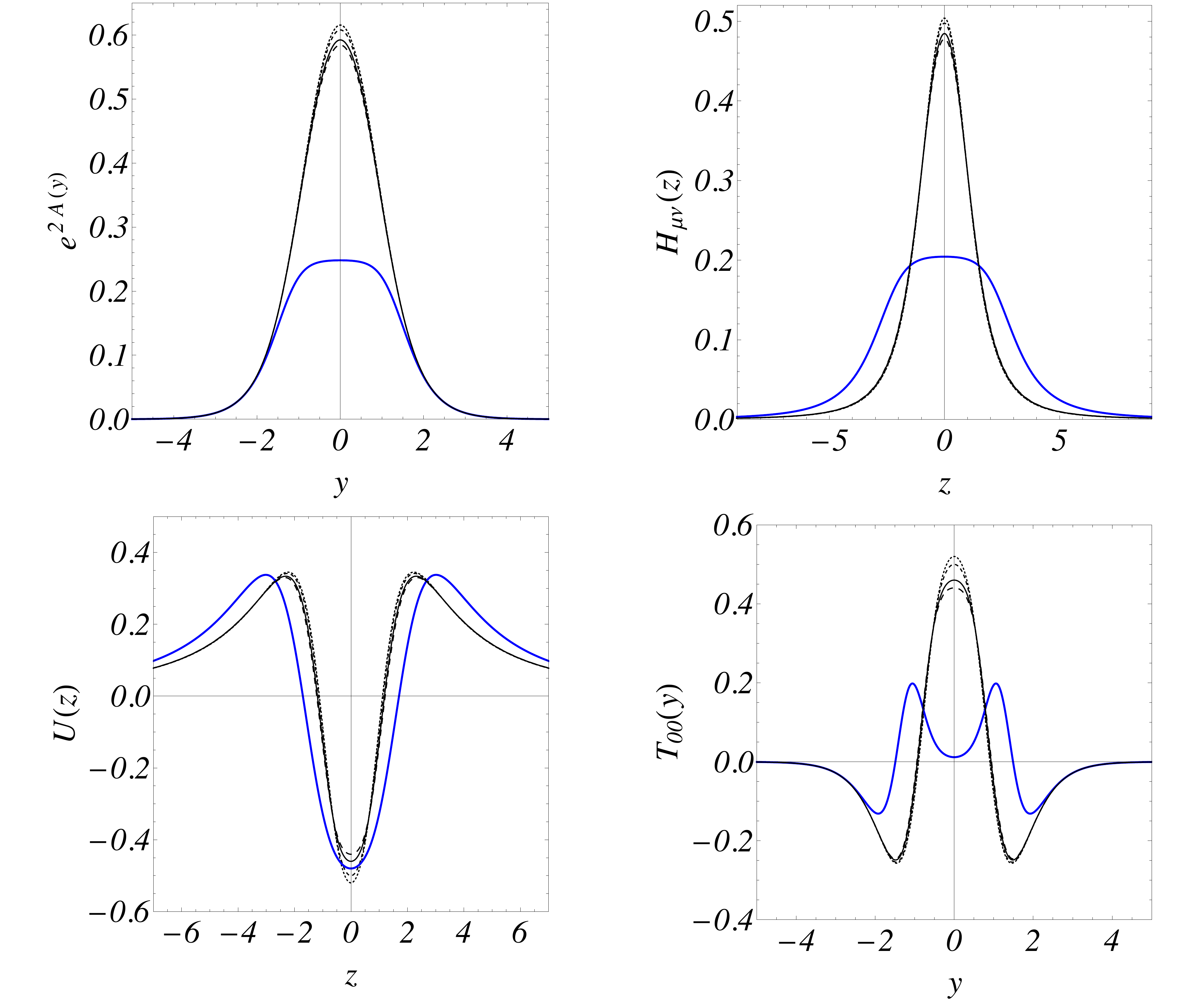}
\caption{Model derived from $\varphi(\chi_3)$ respective to the $4-$cyclic hyperbolic chain. Warp factor ($1^{st}$ column) and zero mode solution ($2^{nd}$ column) are presented in the first line; QM potential as a function of $z$ ($1^{st}$ column) and the energy density ($2^{nd}$ column) are presented in the second line. Blue line represents the solution for $n = 4$ while black lines range from dotted to dashed as $n$ increases (as for the same parameters of Fig.~\ref{lambda1}).}
\label{lambda4.2}
\end{figure}

\begin{figure}[h!]
\centering
\includegraphics[scale=0.4]{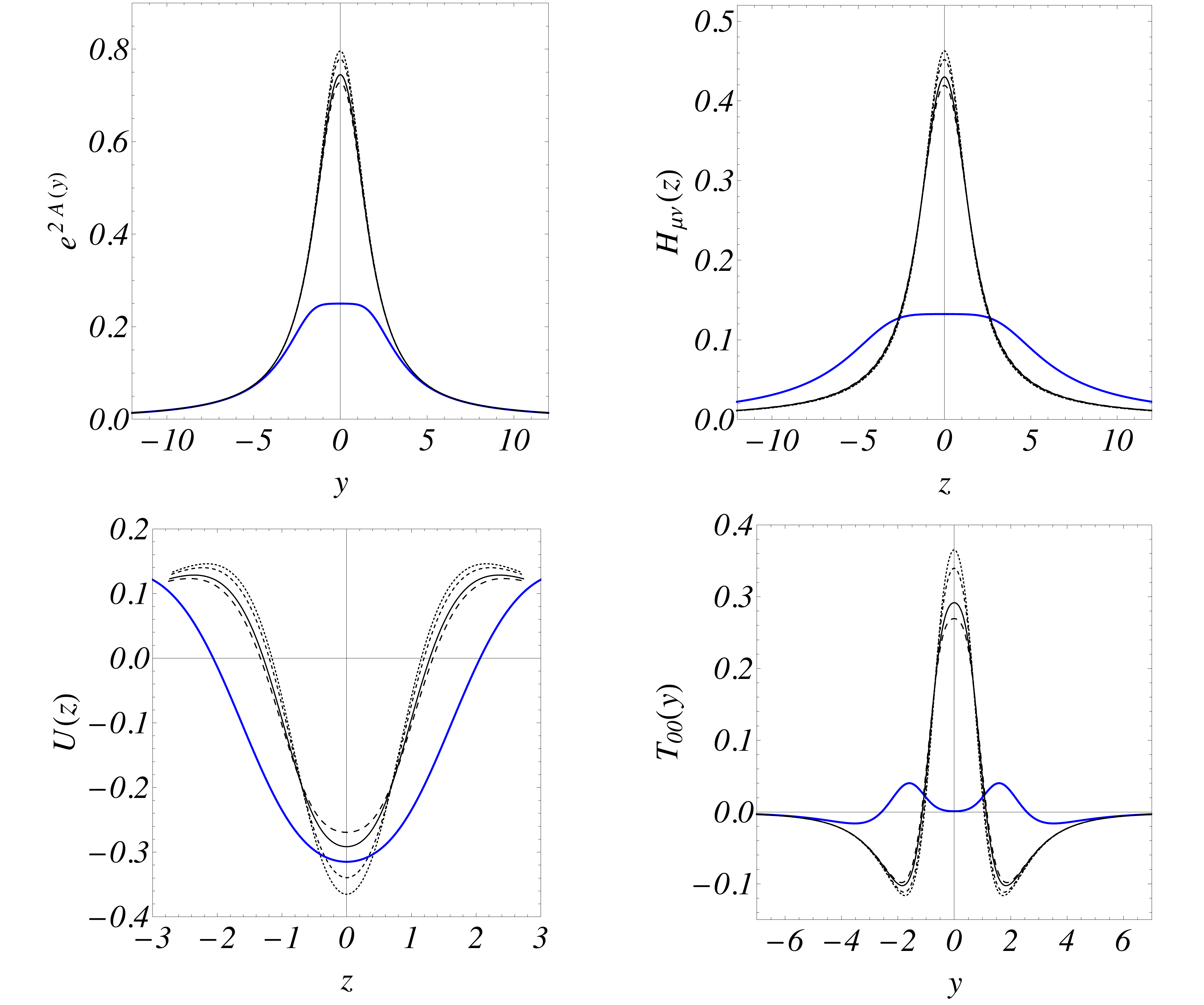}
\caption{model derived from $\varphi(\chi_4)$ respective to the $4-$cyclic hyperbolic chain. Warp factor ($1^{st}$ column) and zero mode solution ($2^{nd}$ column) are presented in the first line; QM potential as a function of $z$ ($1^{st}$ column) and the energy density ($2^{nd}$ column) are presented in the second line. Blue line represents the solution for $n = 4$ while black lines range from dotted to dashed as $n$ increases (as for the same parameters of Fig.~\ref{lambda1}).}
\label{lambda4.3}
\end{figure}

The braneworld scenarios derived from four different deformed lumps (see Table~\ref{tabela}):
\begin{eqnarray}
\phi(\chi_1(y))&=&\frac{\sech[n \tanh(y)] - \sech(n)}{n}, \\
\varphi(\chi_2(y))&=&\frac{1}{n} \tanh[n \sech(y)], \\
\varphi(\chi_3(y))&=& \frac{1}{n} \tanh[n \ln[1 + \sech(y)^2]], \\
\varphi(\chi_4(y))&=& \frac{1}{n} \tanh \bigg[\frac{2n}{(2 + y^2)} \bigg],
\end{eqnarray}
are summarized by the warp factors given by,
\begin{eqnarray}
A_{\phi(\chi 1)}(y) &=& \frac{1}{2} \ln{\left[\frac{\sech[n \tanh(y)]-\sech(n)}{n}\right]},\\
A_{\varphi(\chi 2)}(y) &=& \frac{1}{2} \ln \bigg[\frac{\tanh[n \sech(y)]}{n} \bigg],\\
A_{\varphi(\chi 3)}(y) &=& \frac{1}{2} \ln \bigg[\frac{\tanh[n \log[1 + \sech(y)^2]]}{n} \bigg],\\
A_{\varphi(\chi 4)}(y) &=& \frac{1}{2} \ln \bigg[\frac{1}{n}\tanh\left[\frac{2 n}{2 + y^2}\right] \bigg].
\end{eqnarray}
from which (and from this point) the expressions for the energy density, $T_{00}(y)$ (c. f. Eq.~(\ref{em})), and the corresponding QM potential, $U(z(y))$ (c. f. Eq.~(\ref{QMpotential})) have been suppressed due to their extensive size.
The plots for such complete scenarios derived from the lumps $\phi(\chi_1)$, $\varphi(\chi_2)$, $\varphi(\chi_3)$ and $\varphi(\chi_4)$ are depicted in Figs.~\ref{lambda3}, \ref{lambda4.1}, \ref{lambda4.2} and \ref{lambda4.3} respectively.

The brane scenarios for models depicted in Figs.~\ref{lambda4.2} and \ref{lambda4.3} give rise to thick branes, most of them with no internal structures.
However, for increasing values of the arbitrary parameter $n$ (blue lines), the potential that drives the scalar field allows the appearance of thick branes that host internal structures in the form of a layer of an extra phase circumvented by two separate interfaces.
 
It is related to a more concentrated energy density of the matter field with a corresponding extension/localization of the warp factor which exhibits a profile that approaches to a {\em plateu} form in the region inside the brane.
The mentioned internal structure is noticed from the energy profile. 

\subsubsection{Braneworld for a lump derived from the $4-$cyclic trigonometric chain in a $\lambda \chi^4$ model}

The last braneworld scenario obtained from a deformed lump
\begin{equation}
\varphi(\chi_2(y))=\frac{2 n \sech(y) + \sin[2 n \sech(y)]}{4 n}
\end{equation}
supported by a $\lambda \chi^4$ model leads to the warp factor,
\begin{equation}
A(y) = \frac{1}{2} \ln \bigg[\frac{2 n\, \mbox{sech}(y) + \sin[2 n \,\mbox{sech}(y)]}{4 n} \bigg],
\end{equation}

Now, it is convenient do add some relevant comments to the results obtained up to this point.

By observing the QM potentials obtained for the above solutions supported by $\lambda \chi^4$ model, for all the above obtained solutions, one notices that they exhibit a generalized {\em volcano} behavior and may therefore contain normalized zero modes.
Zero modes were indeed analytically found for all the cases and they are normalizable for the above solutions and for any value of the parameter $n$. It reveals that our brane solutions can support four-dimensional gravity \cite{modelosmundobrana1}.
Another feature that arises from the plot observation is that the variation of the parameter $n$ leads to small changes in the brane profile, changing the values of their critical points.

The plots related to the models derived from $\varphi(\chi_3(y))$ and $\varphi(\chi_4(y))$ were constructed with a different value of $n$, $n=1-2\epsilon;1-\epsilon;1;1+\epsilon;1+2\epsilon$. Thereat, its possible to see the emergence of a \textit{plateau} for $n=4$, opposing to the lower values of $n$, where this structure is not presented.
Furthermore, through the energy-momentum tensor profile, one notices that an interface interpolating two bulk phases is broken into another two separated interfaces. That is, there arises a new phase between the two interfaces, a phenomenon known as complete wetting in condensed matter\cite{completewetting}. Therefore, one can infer that this brane scenario supports internal structures \cite{Brane}. The region where the new phase arises corresponds to the \textit{plateau} region exhibited by the warp factor.

Although the brane defect cannot be analytically calculated, a Taylor series analysis has been implemented as to obtain a robust double-kink approximation for the field solution: this is exactly the kind of defect that one expects in thick brane scenarios, where there internal structures are present \cite{Bas05}.

\begin{figure}[h!]
\centering
\includegraphics[scale=0.4]{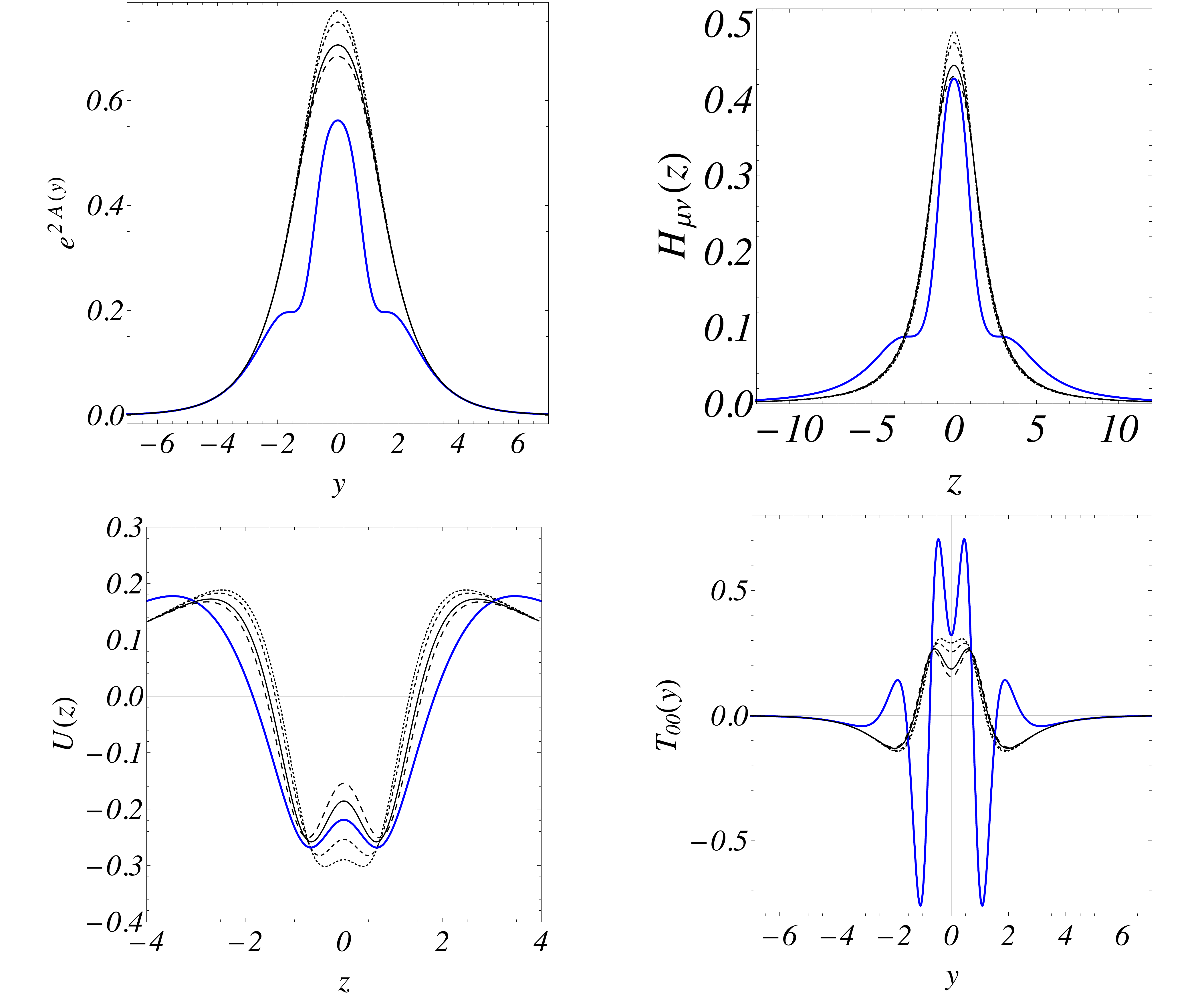}
\caption{Model derived from $\varphi(\chi_2)$ respective to the $4-$cyclic trigonometric chain. Warp factor ($1^{st}$ column) and zero mode solution ($2^{nd}$ column) are presented in the first line; QM potential as a function of $z$ ($1^{st}$ column) and the energy density ($2^{nd}$ column) are presented in the second line. Blue line represents the solution for $n = 4$ while black lines range from dotted to dashed as $n$ increases (as for the same parameters of Fig.~\ref{lambda1}).}
\label{lambda5.1}
\end{figure}

In Fig.~\ref{lambda5.1}, corresponding to the $A_{\varphi(\chi2(y))}$ in the $4-$cyclic hyperbolic chain, one also notices that $n=4$. However, a different behavior arises when one compares it with those results found for $A_{\varphi(\chi3(y))}$ and $A_{\varphi(\chi4(y))}$ relative to the hyperbolic chain.
Through the $e^{2A(y)}$ profile, one notices that there is a lump interpolated by a \emph{plateau}. In this case, through careful observation of the energy-momentum tensor profile, one notices the appearance of several novel phases, generating four matter picks instead of two.
This behavior indicates that the resulting defect has further details in its structure than those presented in the double-kink one.

\section{Conclusions}

Despite its relevance, analytical braneworld scenarios are sparing in the literature. In this work several novel braneworld scenarios have been generated via lump deformed defects inserted in the warp factor calculation. Likewise, the QM analogue problem has been studied by means of metric perturbations. All deformed models are exhibited through an analytical description for which at least one normalizable zero mode has been obtained. In all these cases one therefore can recover $4-$dimensional gravity inside the brane, near $z=0$. In particular, one has noticed the two driving models for thick brane solutions where a double-kink approximation was found for the topological defect. This indicates that the brane supports an internal structure. For the model with a more intricate structure, one sees in the warp factor profile the formation of a lump on the \emph{plateau} structure. For the first time in the literature, a braneworld model supports more than two peaks in the tensor-energy momentum, so that there is the possibility that this brane supports topological defects even more complex than the double-kink, which deserves a more careful investigation.

\section*{Acknowledgement}
{\em The work of AEB is supported by the Brazilian Agency CNPq (grant No. 300809/2013-1 and grant No. 440446/2014-7). RdR is grateful to CNPq (grants No. 303027/2012-6 and No. 473326/2013-2), and to FAPESP (grant 2015/10270-0) for partial financial support.}

\section*{Appendix - Cyclic deformation chain}

A generalized $\kappa$-deformation process, with $\kappa =0,\, 1,\,2,\, \ldots,\,N$  is identified by the $\kappa$-derivative operation given by
\begin{equation}
g^{[\kappa]}(\vartheta^{[\kappa]}) = \frac{d\vartheta^{[\kappa]}}{d\vartheta^{[\kappa-1]}},
\end{equation}
with $\vartheta^{[\kappa]}$ identified by a real scalar fields corresponding to a primitive defect structures constructed from an $N\hspace{-.1 cm}+\hspace{-.1 cm}2$ deformation chain triggered by $\chi\equiv\chi(s)\sim \vartheta^{[-1]}$, such that one effectively has $\vartheta^{[\kappa]} \equiv \vartheta^{[\kappa]}(\chi)$ which can be cyclically recovered by successive deformations.

A hyperbolic chain \cite{lumpalex} is defined through the systematic relations,
\begin{eqnarray}
\vartheta^{[0]}_{\chi} &=& \tanh{(\chi)}, \nonumber\\
\vartheta^{[1]}_{\chi} &=& \tanh{(\chi)}\sech{(\chi)}, \nonumber\\
\vartheta^{[2]}_{\chi} &=& \tanh{(\chi)}\sech{(\chi)}^{2}, \nonumber\\
&\vdots& \nonumber\\
\vartheta^{[N-1]}_{\chi} &=& \tanh{(\chi)}\sech{(\chi)}^{N-1}, \nonumber\\
\vartheta^{[N]}_{\chi} &=& \sech{(\chi)}^{N},
\label{AA}
\end{eqnarray}
where the upper index corresponds to the respective derivative, which upon straightforward integrations results into
\begin{eqnarray}
\vartheta^{[0]}(\chi) &=& \ln{[\cosh{(\chi)}]}, \nonumber\\
\vartheta^{[1]}(\chi) &=& - \sech{(\chi)}, \nonumber\\
\vartheta^{[2]}(\chi) &=& - \frac{1}{2} \sech{(\chi)}^{2}, \nonumber\\
&\vdots& \nonumber\\
\vartheta^{[N-1]}(\chi) &=& - \frac{1}{N-1}\sech{(\chi)}^{N-1}, \nonumber\\
\vartheta^{[N]}(\chi) &=&  \sinh{(\chi)}\, _2F_1\left[\frac{1}{2},\, \frac{1+N}{2},\, \frac{3}{2},\, -\sinh{(\chi)}^2\right], \nonumber\\
\label{BB}
\end{eqnarray}
where $_2F_1$ is the Gauss' hypergeometric function and the integration constants have been suppressed.

From Eqs.~(\ref{AA}-\ref{BB}) one identifies
\begin{eqnarray}
g^{[\kappa]}(\vartheta^{[\kappa]}) = \sech{(\chi)} \equiv \exp{[-\vartheta^{[0]}]},&~~& \kappa = 1,\,2,\, \ldots,\,N-1,\\
g^{[N]}(\vartheta^{[N]}) = 1/\sinh{(\chi)} ,&~~&
\end{eqnarray}
and
\begin{equation}
\prod_{\kappa=1}^{N-1}{g^{[\kappa]}(\vartheta^{[\kappa]})} = \frac{d\vartheta^{[N-1]}}{d\vartheta^{[0]}} = \sech{(\chi)}^{N-1} \equiv \exp{[- (N-1) \vartheta^{[0]}]},
\label{DD}
\end{equation}
which set the complete form of the chain rule of the $N\hspace{-.1 cm}+\hspace{-.1 cm}2$-cyclic deformation, as it can be expressed by
\begin{equation}
\frac{d\vartheta^{[0]}}{d\chi}\,\prod_{\kappa=1}^{N}{g^{[\kappa]}(\vartheta^{[\kappa]})}\,\frac{d\chi}{d\vartheta^{[N]}} = 1,
\label{EE}
\end{equation}
from which a closed cycle of $N\hspace{-.1 cm}+\hspace{-.1 cm}2$ deformation functions is summarized.

Analogously, a trigonometric deformation chain can be engendered by changing $\tanh{(\chi)}\mapsto-\sin{(\chi)}$ and
$\sech{(\chi)}\mapsto\cos{(\chi)}$ into Eqs.~(\ref{AA}-\ref{BB}), as to have
\begin{eqnarray}
\vartheta^{[0]}(\chi) &=& \cos{(\chi)}, \nonumber\\
\vartheta^{[1]}(\chi) &=& \frac{1}{2}\cos{(\chi)}^{2}, \nonumber\\
\vartheta^{[2]}(\chi) &=& \frac{1}{3}\cos{(\chi)}^{3}, \nonumber\\
&\vdots& \nonumber\\
\vartheta^{[N-1]}(\chi) &=& \frac{1}{N}\cos{(\chi)}^{N}, \nonumber\\
\vartheta^{[N]}(\chi) &=&  -\frac{\cos{(\chi)^{N+1}}}{N+1}\, _2F_1\left[\frac{1+N}{2},\, \frac{1}{2},\, \frac{3 + N}{2},\, \cos{(\chi)}^2\right].
\label{BB2}
\end{eqnarray}
as it has been performed in Refs.~\cite{lumpalex,meulump}.

\newpage
\pagebreak

\end{document}